\documentclass{jaa}
\usepackage[square,sort,comma]{natbib}
\usepackage{amsmath}
\usepackage[nopatch]{microtype}
\usepackage{booktabs}
\usepackage{graphicx,subcaption}
\usepackage{amssymb}
\usepackage{lipsum}
\usepackage{float}
\usepackage{hyperref}
\usepackage{rotating} 

\begin{document}
\sloppy

\title{Solar activity and extreme rainfall over Kerala, India}


\author{Elizabeth Thomas\textsuperscript{1}, S. Vineeth\textsuperscript{1} and Noble P. Abraham\textsuperscript{1,*}}
\affilOne{\textsuperscript{1}Department of Physics, Mar Thoma College, Kuttapuzha P. O. Tiruvalla, PIN 689103, Kerala, India.\\}


\twocolumn[{

\maketitle

\corres{noblepa@gmail.com}


\begin{abstract}
This paper examines the relationship between solar activity and extreme rainfall events in Kerala, India. Kerala receives minimum and maximum rainfall during winter and monsoon seasons. Sunspot number, F10.7 Index, and cosmic ray intensity are the solar indices considered, and their variations with rainfall are studied over 57 years (1965-2021), i.e., starting from Solar Cycle 20.
Correlative studies are performed for each solar cycle, and correlation coefficients are calculated. We find that rainfall in Kerala is correlated with sunspot activity but with varying degrees of significance. During Solar Cycle 21, rainfall and solar activity are correlated with high significance during both winter and monsoon seasons. The variation of different solar indices with rainfall is studied. The winter season showed a better link between the sun and rainfall than the monsoon season. The years with rainfall surplus and deficiency are calculated and compared with the solar indices. {We find that the years with rainfall excess and deficit mainly occur in the years around solar maximum or minimum (within $\pm$2 years).}  We hypothesize a physical relationship between solar activity and extreme rainfall events in Kerala that contributes to their predictability.  
\end{abstract}


}]


\doinum{ }
\artcitid{ }
\volnum{ }
\year{2024}
\pgrange{ }
\setcounter{page}{1}
\lp{16}

\section{Introduction}
\label{introduction}
The Earth's climate has been changing drastically over the years. This matter has consistently been a topic of discussion and has not been fully understood. Weather and climate are significantly influenced by the Sun and anthropogenic factors. Solar activity, i.e., magnetic activity inside the Sun, manifests as sunspots, solar flares, solar wind, coronal mass ejections, etc. \citep{Usoskin2017}. Some solar indices commonly used to quantify solar activity are total solar irradiance, sunspot number, solar radio flux, and cosmic rays. The sunspot number quantifies sunspots and is widely used because of its long-term availability. It is strongly correlated with other solar indices \citep{hathaway2015solar,tiwari2018}. A measure of solar radio flux at 10.7 cm is called the F10.7 index, which originates deep in the corona and high in the chromosphere \citep{tapping1994limits,tapping201310}. Cosmic rays originate outside the solar system and are high-energy particles reaching Earth. It is observed that cosmic rays are negatively correlated with sunspot number \citep{Gupta2006}.

Recently, there has been an increased interest in studying solar-climate relations, and several studies have indicated that changes in solar activity could affect regional climates \citep{rycroft, Tinsley, Stefani}. Solar activity affects rainfall in a wide variety of ways, which are reflected in the varying correlations based on the time scale and region \citep{Tsiropoula2003, ZhaoJuanHanYan-BenandLi2004,Wasko2009,Mauas2011,Rampelotto2012}. Some recent work on sun-rainfall link were carried out in China \citep{Zhai2017,YU2019,yan2022}, the United States \citep{Nitka2019}, Europe \citep{Laurenz2019}, Argentina \citep{HEREDIA2019105094}, Nepal \citep{tiwari2021} and Northeast Asia \citep{yan2022}. 

Changes in rainfall patterns can severely affect a country like India, posing challenges to its agriculture, food security, economy, and ecosystem \citep{DoranaluChandrashekar2017}. Over India as a whole or in different regions, numerous authors have explored the possibility of an association between solar activity and rainfall \citep{jagannathan1973changes,Ananthakrishnan1984,Hiremath2004,Bhattacharyya2005,Agnihotri2011,Badruddin2015,Warrier2017,Thomas2022}. The direct and indirect effects were studied. The results were often localised and contradicted other authors \citep{JAGANNATHAN1973,bhalme1981solar,Hiremath2006,Bhattacharyya2007,Lihua2007,Selvaraj2009,selvaraj2011study,Selvaraj2013,Hiremath2015,Malik2018,Thomas2023}.

\begin{figure*}[t]
\centering
\includegraphics[scale=0.45]{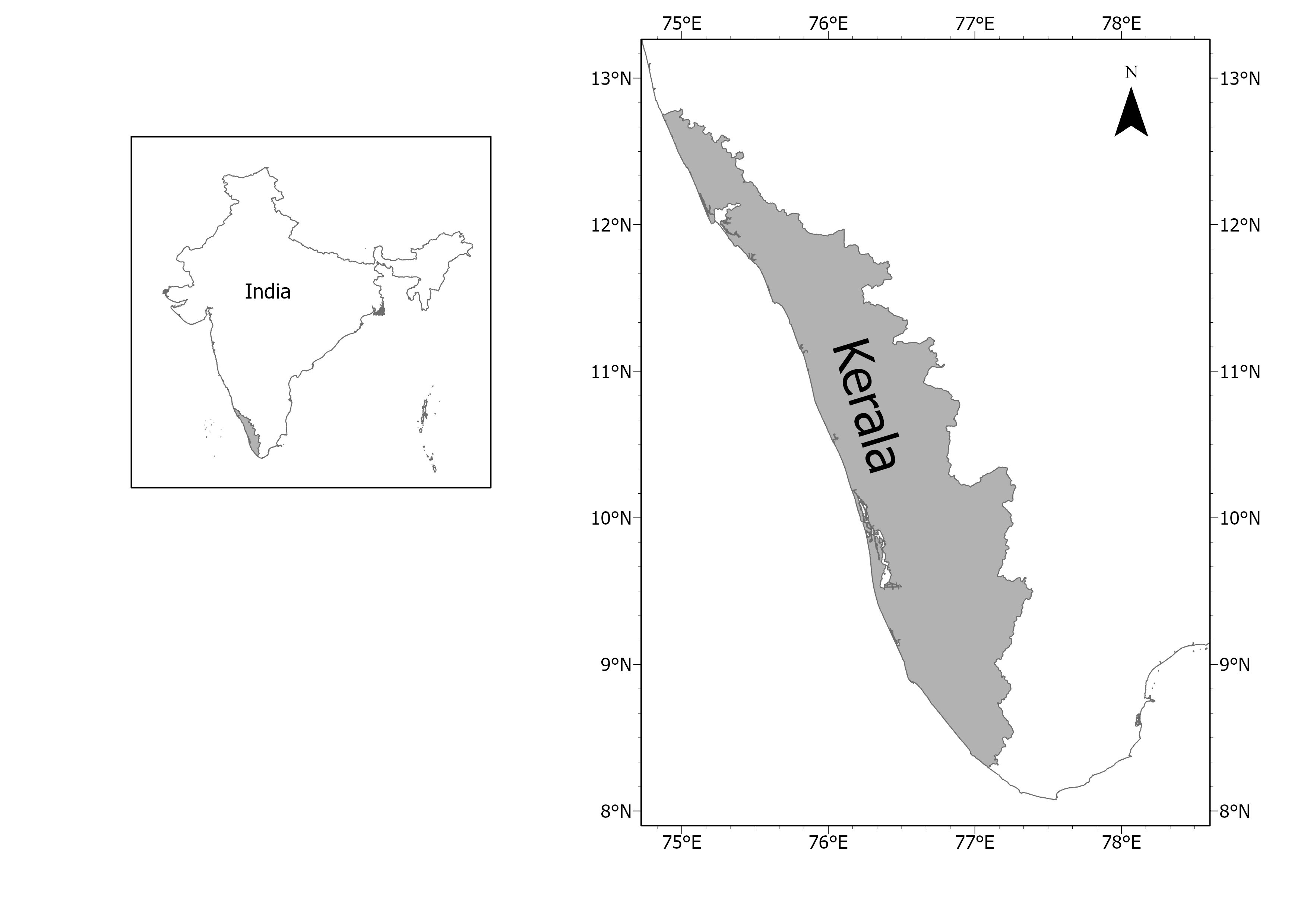}
\caption{Location map of Kerala}
\label{kerala}
\end{figure*}

Kerala lies on the southwestern tip of India and is bounded east by the Western Ghats and west by the Arabian Sea. It extends between 8$^{\circ}$15$^{\prime}$ and 12$^{\circ}$50$^{\prime}$ northern latitude and between 74$^{\circ}$50$^{\prime}$ and 77$^{\circ}$30$^{\prime}$ eastern longitude. The climate of Kerala is subtropical, with the eastern highlands (rugged and cool mountainous), the central midlands (rolling hills), and the western lowlands (coastal plains). Kerala's diverse features make it more vulnerable to climate change. Kerala is known as the "gateway to the summer monsoon." Studies on long-term rainfall variability found that southwest monsoon rainfall decreased significantly while post-monsoon rainfall increased \citep{Krishnakumar2009,Kothawale2017}. Recently, a few studies have reported the influence of sunspot number on rainfall over Kerala \citep{thomas2022impact,Thomas2022,Thomas2023}. The location map of Kerala is shown in Figure \ref{kerala}.

\begin{figure*}[t]
\centering
\includegraphics[scale=0.45]{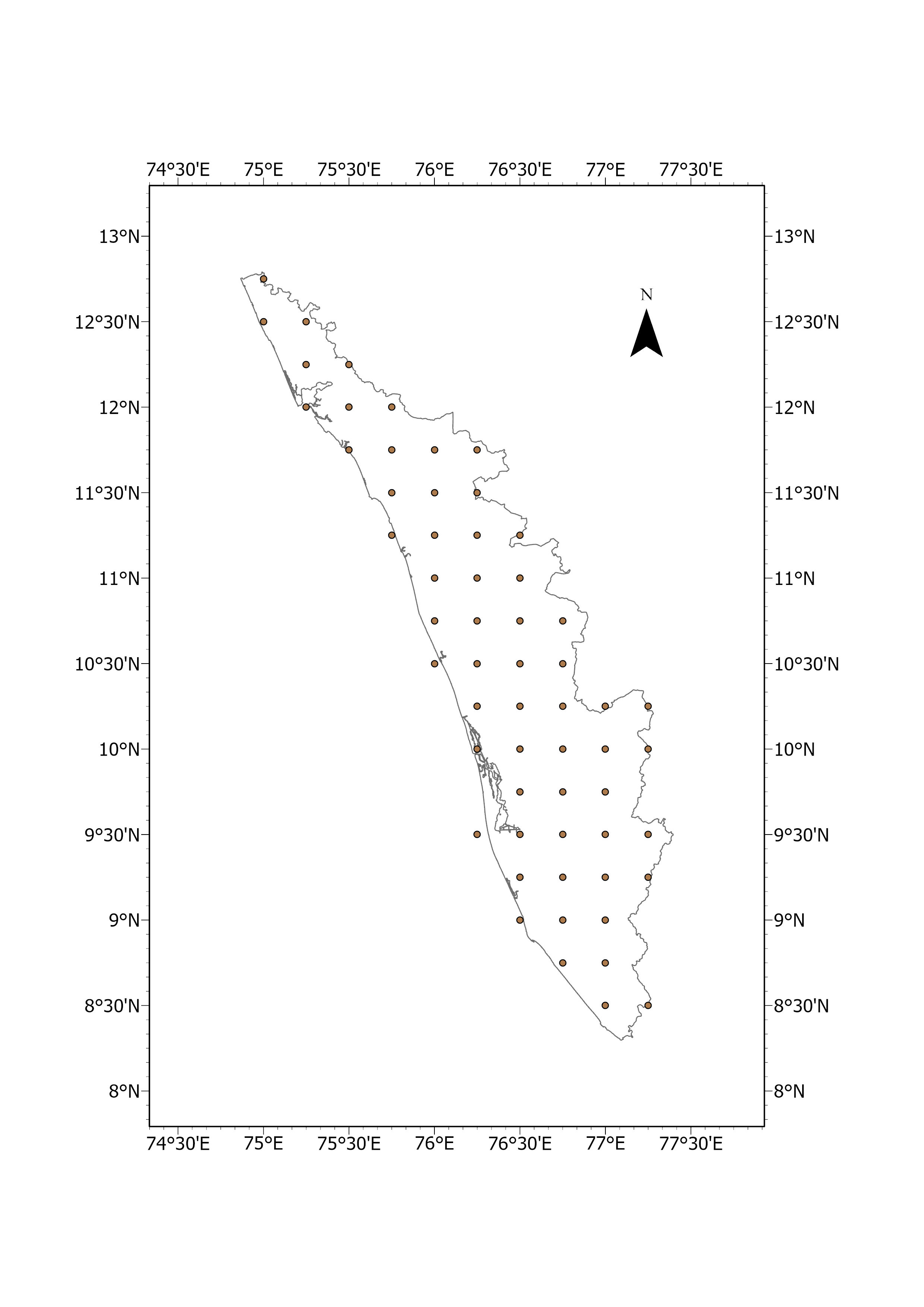}
\caption{Location of Grid points}
\label{keralagrid}
\end{figure*}

Extreme rainfall events have recently occurred in Kerala, causing floods or landslides that have resulted in the loss of lives and property. In India, several studies have linked solar activity to extreme weather events (see, e.g., \cite{bhalme1981cyclic,Azad2011}). Hence, studying extreme rainfall over the Kerala region with different solar parameters would be interesting. 

Investigating the response of rainfall to different solar indices could reveal subtle differences. This paper evaluates the influence of different solar indices, i.e., sunspot number, F10.7 Index, and cosmic ray intensity, on extreme precipitation over Kerala using correlation studies.
Section \ref{data} discusses the data and methodology of the analysis. Section \ref{results} presents the results and discussion on the correlation and variation of different solar indices with rainfall over Kerala. It also includes results about extreme rainfall events. Section \ref{conclusions} contains the conclusions.

\section{Data and Methods}\label{data}

\subsection{Dataset} \label{dataset}

This study used daily data of various solar indices (sunspot number, F10.7 Index, and cosmic ray intensity) and rainfall over Kerala from 1965 to 2021, i.e., from Solar Cycle 20. The sunspot number data was obtained from the World Data Center SILSO, Royal Observatory of Belgium, Brussels. F10.7 Index is the solar flux data (in sfu, 1 sfu = $10^{-22} W m^{-2} Hz^{-1}$) and was downloaded from LASP Interactive Solar Irradiance Data Center. Cosmic ray intensity (in counts/min) data was taken from the Oulu Cosmic Ray station. Rainfall (in mm) over Kerala was obtained from the India Meteorological Department’s (IMD) daily gridded rainfall dataset of high spatial resolution (0.25$^{\circ}$ × 0.25$^{\circ}$) \citep{Pai2014}. 59 grid points covering the region of Kerala were used in this study and are shown in Figure \ref{keralagrid}.

India Meteorological Department (IMD) classifies the seasons of India as Winter (January-February), Pre-monsoon (March-May), Southwest monsoon (June-September), and Post-monsoon (October-December).

\begin{figure*}[ht]
\centering
  \begin{subfigure}{0.5\textwidth}
    \includegraphics[width=\linewidth]{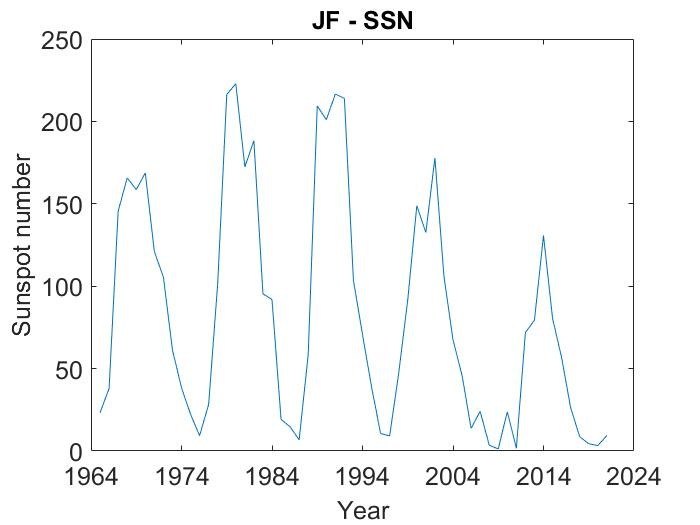}
     \label{fig:3a}
  \end{subfigure}%
  \begin{subfigure}{0.5\textwidth}
    \includegraphics[width=\linewidth]{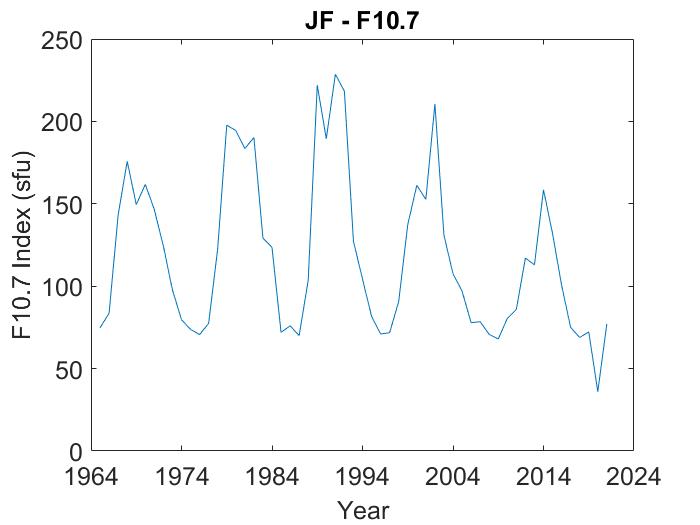}
     \label{fig:3b}
  \end{subfigure}  \\
   \begin{subfigure}{0.5\textwidth}
    \includegraphics[width=\linewidth]{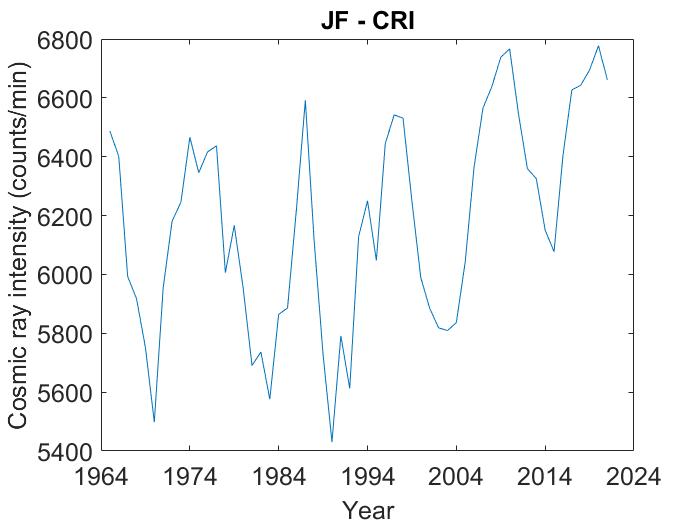}
     \label{fig:3c}
  \end{subfigure}%
  \begin{subfigure}{0.5\textwidth}
    \includegraphics[width=\linewidth]{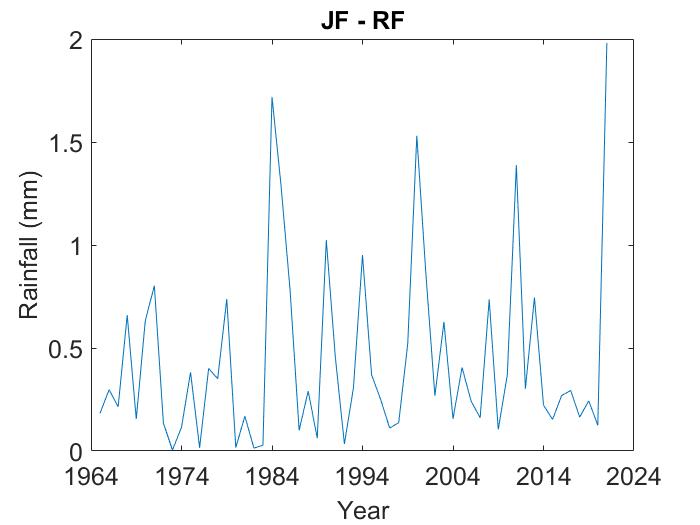}
     \label{fig:3d}
  \end{subfigure}  \\
\caption{Time series of (a) sunspot number (SSN) (b) F10.7 index (F10.7) (c) cosmic ray intensity (CRI) and (d) rainfall (RF), corresponding to JF season.}
\label{jfseason}
\end{figure*}

\begin{figure*}[ht]
\centering
  \begin{subfigure}{0.5\textwidth}
    \includegraphics[width=\linewidth]{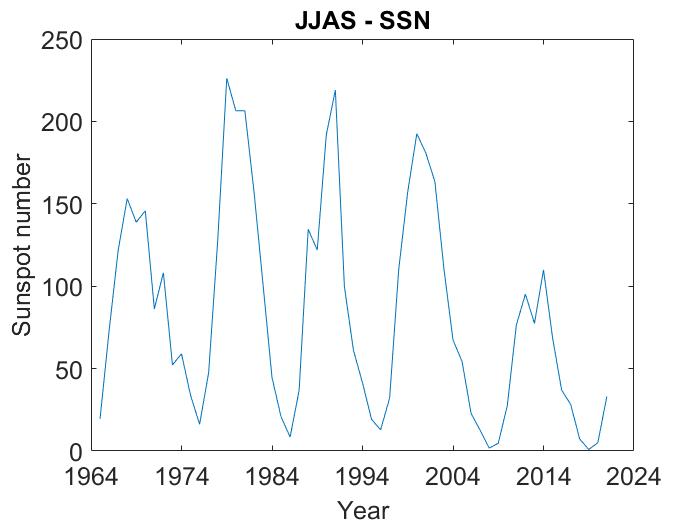}
     \label{fig:4a}
  \end{subfigure}%
  \begin{subfigure}{0.5\textwidth}
    \includegraphics[width=\linewidth]{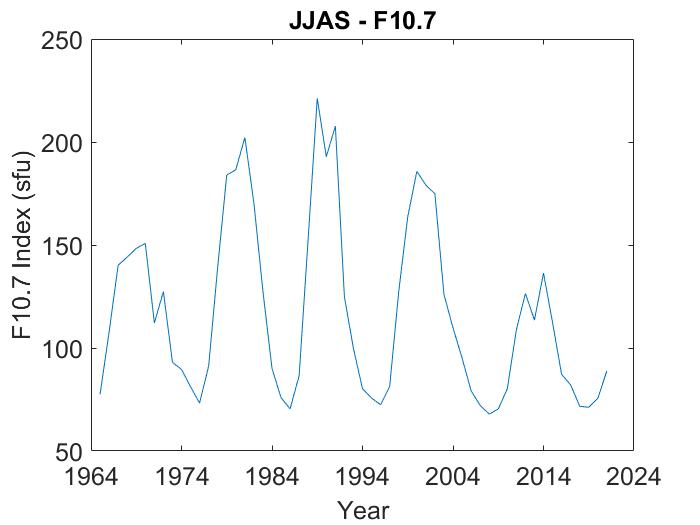}
     \label{fig:4b}
  \end{subfigure}  \\
   \begin{subfigure}{0.5\textwidth}
    \includegraphics[width=\linewidth]{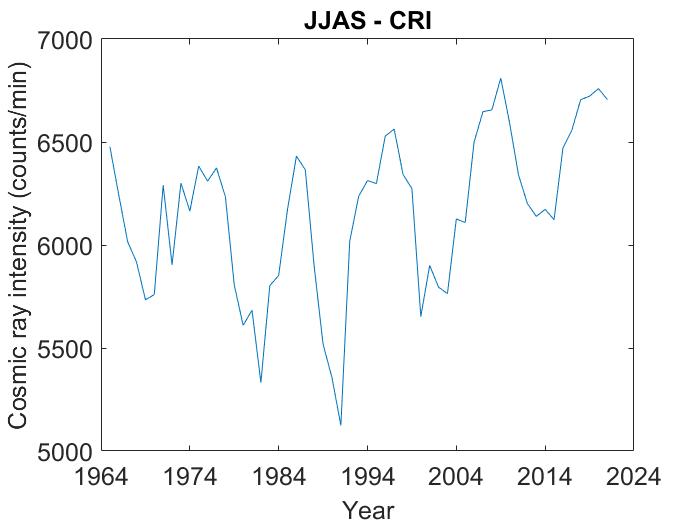}
     \label{fig:4c}
  \end{subfigure}%
  \begin{subfigure}{0.5\textwidth}
    \includegraphics[width=\linewidth]{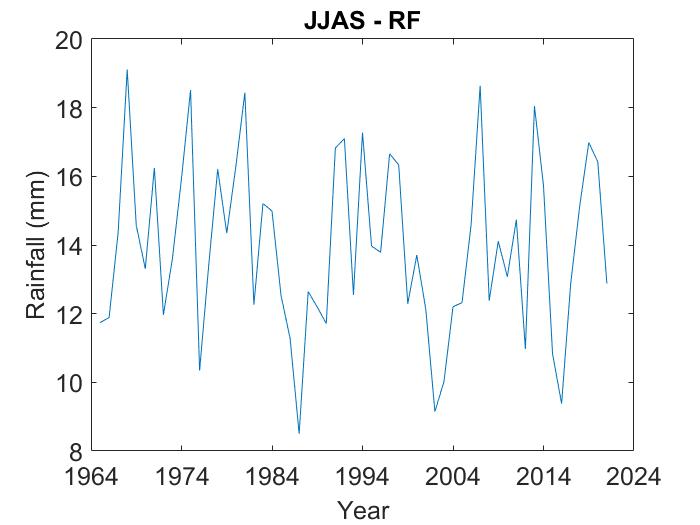}
    \label{fig:4d}
  \end{subfigure}  \\
\caption{Time series of (a) sunspot number (SSN) (b) F10.7 index (F10.7) (c) cosmic ray intensity (CRI) and (d) rainfall (RF), corresponding to JJAS season.}
\label{jjasseason}
\end{figure*}

This study used the Southwest-monsoon season, denoted as JJAS, and the winter season, denoted as JF, as maximum and minimum rainfall are accounted for during these seasons. A similar grouping (JF and JJAS) was done with the solar activity features (sunspot number, F10.7 index, and cosmic ray intensity) data. The daily data for each parameter were averaged for each season, and the corresponding values for sunspot number are given as SSN (/days), F10.7 Index is given as F10.7 (sfu/days), cosmic ray intensity is given as CRI (count/min/days), and rainfall is given as RF (mm/days). Figure \ref{jfseason} and Figure \ref{jjasseason} show the time series of SSN, F10.7, CRI, and RF corresponding to the JF and JJAS seasons, respectively.

\subsection{Methodology} \label{methodology}
Correlative studies were carried out to observe the relationship between solar indices and rainfall for Solar Cycles 20-24.
Correlation coefficients are usually computed to check whether any relationship exists between the two data sets and how strong the relationship is. Here, Spearman Rank-Order correlation coefficients and their significance were calculated to determine the relationship between different solar indices (SSN, F10.7, and CRI) and rainfall (RF) data. Since this correlation method provides information about monotonic relationships, it is more powerful than linear correlation \citep{Hiremath2004,Hiremath2006,Bankoti2011}. 
The p-value obtained gives the significance of the test. The p-values $<$ 0.1 gives 10 \% significance. Corresponding to all the solar activity indices, correlation coefficients and their significance were determined for each solar cycle. As the rainfall is noisy, the RF values were smoothened by moving averages of two, four, and six points, and the correlation studies were carried out \citep{Bankoti2011}. 

\par

In order to study the relation of extreme rainfall events with solar activity, the years of excess and deficit rainfall over Kerala were identified. For that, the mean ($\mu$) and standard deviation ($\sigma$) of rainfall RF during both seasons (JF and JJAS) were determined. A year $i$ was labelled as extreme rainfall year when $R_{i} \geq (\mu + \sigma)$ and a year labelled as deficient rainfall year when 
$R_{i} \leq (\mu - \sigma)$, where $R_{i}$ is the rainfall of that year, $ i,k\in\mathbb{R} $ \citep{Azad2011}. In this study, $k$ was defined as one. The excess and deficient rainfall years were used to study the relation of rainfall with solar activity.

\section{Results and Discussions}\label{results}

The Spearman Rank-Order correlation coefficient and its significance between different solar indices, i.e., sunspot number, F10.7 Index and, cosmic ray intensity, and rainfall, is shown in Table \ref{corr}, where * represents 0.1 significance level. The first column represents the solar cycle; the second column represents different types of moving point averages and original data sets for different seasonal months (JF and JJAS). The correlation coefficients with significance(in brackets) are represented in columns three, four, and five of SSN, F10.7, and CRI with RF, respectively.

\begin{table*}[t]
 \centering
  \caption{The correlation coefficients and significance of correlation (in brackets) between different solar indices (SSN, F10.7 and CRI) and rainfall RF for the seasonal months (JF and JJAS). Correlation coefficients for the same solar activity indices with two, four and six point moving averages of rainfall are also represented.}
     \label{corr}
    \begin{tabular}{ccccc} \cline{1-5}
       Solar Cycle & Seasonal months & \multicolumn{3}{c}{Solar activity indices }\\ \cline{1-5}
       & & SSN & F10.7 & CRI \\\cline{3-5}
      20 & JF(original) & 0.43 (0.16) & 0.51 (0.09)* & -0.50 (0.10)* \\
         & 2 pt & 0.61 (0.04)* & 0.69 (0.02)* & -0.76 (0.01)* \\
         & 4 pt & 0.83 (0.001)* & 0.83 (0.001)* & -0.83 (0.002)* \\
         & 6 pt & 0.85 (0.001)* & 0.89 (0.0001)* & -0.83 (0.001)* \\
         & JJAS(original) & 0.43 (0.16) & 0.34 (0.28) & -0.21 (0.51) \\
         & 2 pt & 0.24 (0.46) & 0.17 (0.59) & -0.23 (0.47) \\
         & 4 pt & 0.39 (0.20) & 0.41 (0.19) & -0.51 (0.09) \\
         & 6 pt & 0.19 (0.54) & 0.34 (0.29) & -0.35 (0.27) \\ \hline
         21 & JF(original) & -0.66 (0.04)* & -0.57 (0.09)* & 0.38 (0.28) \\
         & 2 pt & -0.54 (0.11) & -0.51 (0.13) & 0.57 (0.09)* \\
         & 4 pt & -0.75 (0.02)* & -0.71 (0.03)* & 0.30 (0.41) \\
         & 6 pt & -0.87 (0.003)* & -0.76 (0.02)* & 0.15 (0.68) \\
         & JJAS(original) & 0.56 (0.09)* & 0.68 (0.03)* & -0.39 (0.26) \\
         & 2 pt & 0.76 (0.01)* & 0.83 (0.006)* & -0.83 (0.006)* \\
         & 4 pt & 0.82 (0.007)* & 0.87 (0.003)* & -0.88 (0.002)* \\
         & 6 pt & 0.95 (0)* & 0.93 (0.0001)* & -0.66 (0.04)* \\ \hline
         22 & JF(original) & 0.06 (0.86) & 0.08 (0.3) & -0.09 (0.81) \\
         & 2 pt & 0.2 (0.43) & 0.26 (0.47) & -0.31 (0.39) \\
         & 4 pt & -0.07 (0.86) & -0.11 (0.76) & -0.15 (0.68) \\
         & 6 pt & 0.16 (0.65) & 0.13 (0.73) & 0.05 (0.89) \\ 
         & JJAS(original) & -0.08 (0.84) & -0.30 (0.47) & 0.06 (0.86) \\
         & 2 pt & -0.22 (0.53) & -0.30 (0.41) & 0.09 (0.81) \\
         & 4 pt & -0.25 (0.49) & -0.38 (0.28) & 0.23 (0.51) \\
         & 6 pt & -0.37 (0.29) & -0.50 (0.14) & 0.27 (0.45) \\  \hline
         23 & JF(original) & 0.44 (0.15) & 0.47 (0.13) & -0.30 (0.34) \\
         & 2 pt & 0.65 (0.03)* & 0.65 (0.03)* & -0.56 (0.06)* \\ 
         & 4 pt & 0.81 (0.002)* & 0.82 (0.002)* & -0.54 (0.07)* \\
         & 6 pt & 0.83 (0.001)* & 0.78 (0.004)* & -0.68 (0.02)* \\
         & JJAS(original) & -0.56 (0.06)* & -0.51 (0.09)* & 0.67 (0.02)* \\
         & 2 pt & -0.50 (0.09)* & -0.44 (0.15) & 0.79 (0.004)* \\
         & 4 pt & -0.48 (0.12) & -0.42 (0.17) & 0.76 (0.006)* \\ 
         & 6 pt & -0.42 (0.17) & -0.36 (0.26) & 0.69 (0.001)* \\ \hline  
         24 & JF(original) & -0.02 (0.97) & 0.14 (0.70) & 0.03 (0.94) \\
         & 2 pt & 0.16 (0.65) & 0.34 (0.33) & -0.22 (0.54) \\
         & 4 pt & 0.27 (0.44) & 0.47 (0.18) & -0.25 (0.49) \\
         & 6 pt & -0.33 (0.35) & -0.07 (0.83) & 0.21 (0.56) \\
         & JJAS(original) & 0.16 (0.65) & 0.10 (0.78) & -0.001 (1) \\
         & 2 pt & 0.15 (0.68) & 0.13 (0.73) & -0.07 (0.86) \\
         & 4 pt & 0.44 (0.20) & 0.45 (0.19) & -0.36 (0.31) \\
         & 6 pt & 0.42 (0.23) & 0.34 (0.33) & -0.18 (0.63) \\ \hline
        
    \end{tabular}
     
     Note. * indicates higher than 0.1 significance level.
   \end{table*}

\subsection{Relationship between Sunspot number and Rainfall}

\begin{figure*}[ht]
\centering
  \begin{subfigure}{0.5\textwidth}
    \includegraphics[width=\linewidth]{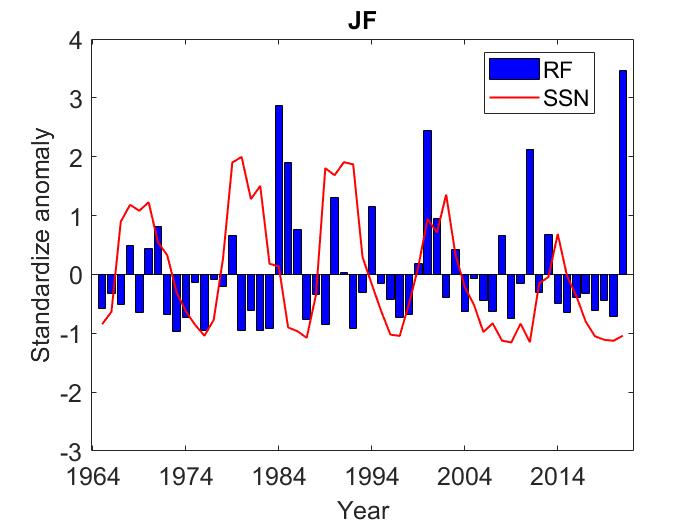}
     \label{fig:5a}
  \end{subfigure}%
  \begin{subfigure}{0.5\textwidth}
    \includegraphics[width=\linewidth]{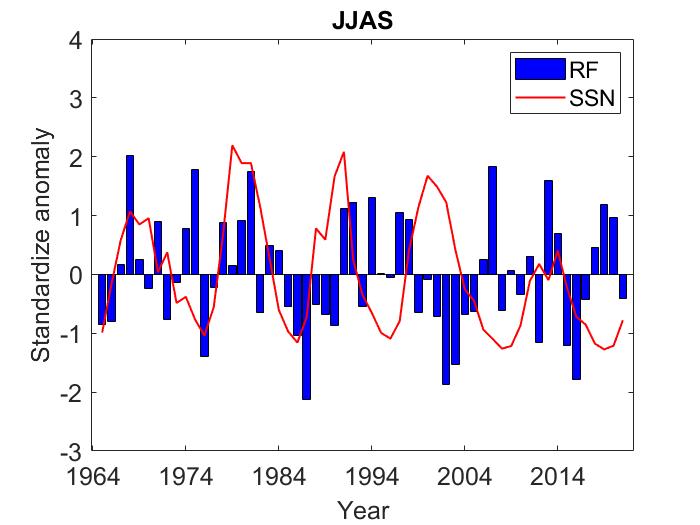}
     \label{fig:5b}
  \end{subfigure}  \\
\caption{Variation of sunspot number (SSN) and rainfall (RF) during (a) JF season and (b) JJAS season.}
\label{ssnrf}
\end{figure*} 
 {Data standardization was done as the dataset was collected from different locations. The standardized anomaly was calculated by subtracting the mean and dividing it by its standard deviation for both seasons. The variation in standardized anomaly represents a deviation from the mean state. The anomaly is positive when the data is increasing and negative when decreasing. For easy comparison between the data, the variation in solar indices was represented as a line graph and the rainfall as a bar graph, with each bar representing a year.  }
Figure \ref{ssnrf} represents the time series of standardized values of sunspot number (SSN) and rainfall (RF) during the JF and JJAS seasons. Solar cycles 20-24 were covered in this study. To study the response of rainfall over Kerala to the sunspot number, the Spearman rank-order correlation coefficients between SSN and RF were calculated. First, the entire 57 years were considered, and the correlation coefficients were found to be low, i.e., 0.03 and  0.002 during the JF and the JJAS seasons, respectively. Solar cycle-wise correlation coefficients were then determined, and the results are given in Table \ref{corr}. The rainfall over Kerala was noted to be correlated with sunspot number, with varying significance, irrespective of signs.

For the JF season, significant correlations, irrespective of signs, were observed during Solar cycles 20, 21, and 23. The correlation coefficients were noted to be negative only during Solar cycle 21, positive during Solar cycles 20 and 23, and changed signs on smoothing RF values during Solar cycle 22 and 24. It was seen that the significance of the correlation coefficients improved in smoothing the rainfall RF values by two, four, and six-point moving averages. {Among the five solar cycles considered, Solar cycles 22 and 24 showed a weak correlation.} For the JJAS season, Solar Cycle 21 showed the highest correlation between SSN and RF values with significance for original, two, four, and six-point moving average values. The correlation coefficients were positive during Solar cycles 20, 21, and 24 and negative during Solar cycles 22 and 23. Among the two seasons, the JF season revealed a better association between sunspot numbers and rainfall over Kerala. 

\subsection{Relationship between F10.7 Index and Rainfall}

\begin{figure*}[ht]
\centering
  \begin{subfigure}{0.5\textwidth}
    \includegraphics[width=\linewidth]{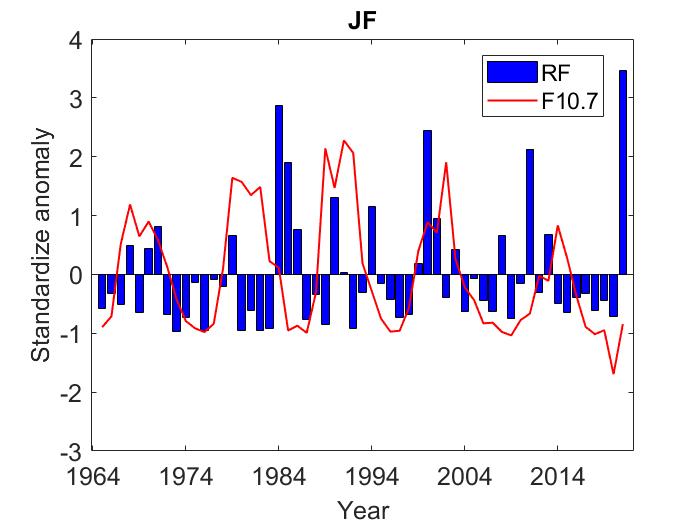}
     \label{fig:6a}
  \end{subfigure}%
  \begin{subfigure}{0.5\textwidth}
    \includegraphics[width=\linewidth]{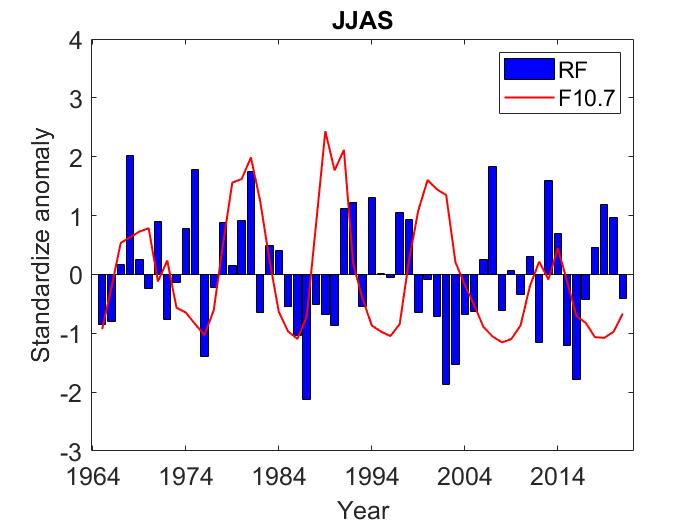}
  \label{fig:6b}
  \end{subfigure}  \\
\caption{Time series of F10.7 Index (F10.7) and rainfall (RF) during (a) JF season and (b) JJAS season.}
\label{f10.7rf}
\end{figure*} 

Figure \ref{f10.7rf} shows the time series of standardized values of the F10.7 Index (F10.7) and rainfall (RF) during the JF and JJAS seasons. Here, the correlation coefficient was also computed to be 0.09 and -0.03 during the JF and the JJAS seasons, respectively, considering the entire period of study. Correlation coefficients corresponding to Solar cycles 20-24 were worked out, and the results are given in Table \ref{corr}. Like SSN results, the rainfall values were observed to be correlated with the F10.7 Index with varying significance, regardless of the signs. During the JF season, Solar cycles 20, 21, and 23 showed a high correlation with significance. The correlation appeared to improve in smoothing the rainfall values. In the case of the JJAS season, Solar Cycle 21 revealed a highly significant correlation compared to other cycles. {It can be concluded that rainfall in Kerala is related to solar activity during the JF season since the F10.7 Index showed a high correlation during three solar cycles during this season compared to the JJAS season.}

\subsection{Relationship between Cosmic ray intensity and Rainfall}

\begin{figure*}[ht]
\centering
  \begin{subfigure}{0.5\textwidth}
    \includegraphics[width=\linewidth]{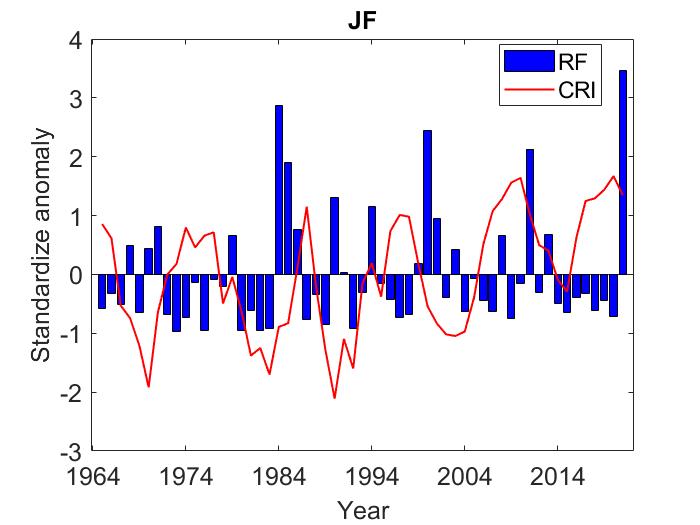}
     \label{fig:7a}
  \end{subfigure}%
  \begin{subfigure}{0.5\textwidth}
    \includegraphics[width=\linewidth]{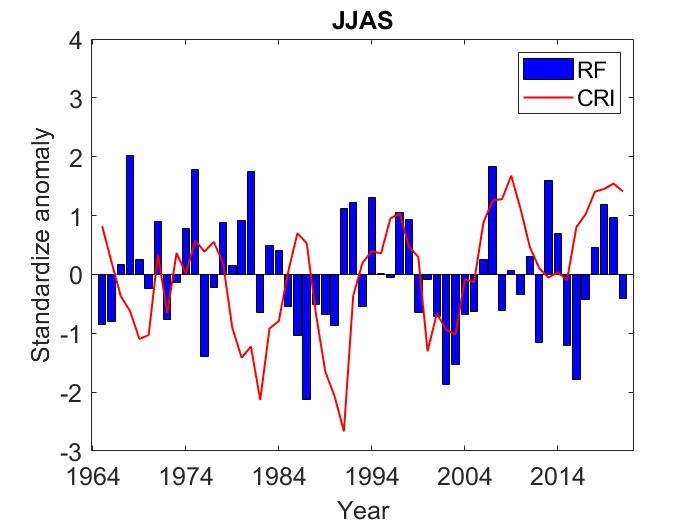}
   \label{fig:7b}
  \end{subfigure}  \\
\caption{Time series of Cosmic ray intensity (CRI) and rainfall (RF) during (a) JF season and (b) JJAS season.}
\label{crirf}
\end{figure*} 

Figure \ref{crirf} shows the time series of standardized values of cosmic ray intensity (CRI) and rainfall (RF) during the JF and the JJAS seasons. During the JF season, it was observed that the rainfall anomaly is negative during the decreasing phase of CRI in the first solar cycle starting from 1965. In the next cycle, starting from 1977, the rainfall anomaly is observed to be negative during the decreasing phase of CRI again. Out of the five decreasing phases observed, the first three rainfall anomalies were negative, and the remaining two were positive. In the case of the JJAS season, the rainfall anomalies alternate during each decreasing phase of CRI. During the first cycle, rainfall anomaly is negative and during the next cycle, it is observed to be positive, and so on. \cite{Chaudhuri2015} reported results showing that the decreasing phase of GCR is crucial for identifying rainfall variability.

Correlation coefficients computed between CRI and RF values considering the entire period were found to be -0.08 and 0.10 during the JF and the JJAS seasons, respectively. Correlation coefficients were determined for each solar cycle, and the results are given in Table \ref{corr}. {Compared to sunspot number and F10.7 Index, cosmic ray intensity showed correlation results often opposite in sign.} Solar cycles 20 and 23 showed a significant negative correlation between CRI and RF during the JF season, whose significance improved in smoothing the rainfall data. Solar Cycles 21 and 23 revealed a significant correlation during the JJAS season for smoothed rainfall values. 

Solar Cycle 21 (1977-1986) was the strongest of the five solar cycles considered, and Solar Cycle 24 (2009-2018) was the weakest in terms of SSN. It was noted that when the solar activity was maximum, a high correlation with significance was observed in the JF season. {A weak correlation between sunspot number and rainfall was found during periods of low solar activity, i.e., during Solar Cycle 24.} Solar cycles 20, 21, and 23 showed a good association between SSN and RF, and after smoothing the data, the correlation improved. As for the JJAS season, Solar Cycle 21 showed a positive correlation, but this correlation decreased during Solar Cycle 24 when solar activity was lower. These results are consistent with the earlier results. \cite{Thomas2022} analyzed the relationship between sunspot number and rainfall over Kerala during varying levels of solar activity using wavelet coherence and noted higher coherence during the high solar activity period than during low solar activity, during winter (JF) and monsoon (JJAS) seasons. 
Considering the F10.7 Index and Cosmic ray intensity parameters, the results were similar, and the correlation results were weaker than those of the sunspot number. When comparing the two seasons, the JF season had a more decisive solar influence on its rainfall than the JJAS season.

Several correlative studies have been conducted in India to determine if solar influences affect rainfall there. \cite{Ananthakrishnan1984} reported both positive and negative correlation coefficients while considering 306 stations in India. \cite{Hiremath2004} investigated the correlative effects of sunspot number on the seasonal and annual Indian monsoon rainfall and found that the pre-monsoon and monsoon rainfall showed significant positive correlations. \cite{Hiremath2006} analyzed correlative effects of sunspot number over the Indian rainfall corresponding to each solar cycle and noted correlation irrespective of the signs, with a moderate to high significance. 
When solar activity was low, rainfall was higher than when solar activity was high. \cite{Bal2010} reported the existence of weak positive and negative correlations during different seasons. \cite{Bankoti2011} conducted several statistical studies between the solar parameters (Sunspot number, solar active prominences, and H alpha solar flares) and All India homogeneous rainfall and noted that the correlation varied its sign with different seasons and also with different solar parameters. \cite{Chaudhuri2015} performed seasonal correlation and observed a possible association between cosmic rays and rainfall during the post-monsoon season. Several correlative studies were carried out in different states to find a possible sun-rainfall link, i.e., in West Bengal \citep{chakraborty1986solar}, Rajasthan \citep{jain1997correlation}, Tamil Nadu \citep{Selvaraj2009,selvaraj2011study,selvaraj2012} and Kerala \citep{thomas2022impact}.

\subsection{Solar activity indices and extreme rainfall in Kerala}

A study was conducted to examine the possible relationship between solar activity and extreme rainfall events in Kerala during the JF and JJAS seasons. The years of excess and deficient rainfall were identified, as explained in Section \ref{methodology} \citep{Azad2011}. During the JF season, six excess rainfall events were visible in 1984, 1990, 1994, 2000, 2011, and 2021. This season was not marked by deficient rainfall. In the case of the JJAS season, both excess and deficient rainfall events were observable. Ten years of excess rainfall were recorded in 1968, 1975, 1981, 1991, 1992, 1994, 1997, 2007, 2013 and 2019. Similarly, five years of deficient rainfall were observed during 1976, 1987, 2002, 2012, and 2016. 

A study of the relative timing of solar activity and extreme rainfall was conducted using curves of different solar activity features corresponding to different seasons. Figure \ref{ssnexcess}, \ref{f107excess}, and \ref{criexcess} represent the extreme rainfall events with SSN, F10.7, and CRI, respectively. The black circles denote excess rainfall years, and the red circle denotes deficient rainfall years. The present study covers five complete solar cycles, i.e., Solar Cycles 20-24. Tables \ref{ssnextremejf}, \ref{f10.7extremejf} and \ref{criextremejf} list the extreme rainfall events during the JF season, and Tables \ref{ssnextremejjas}, \ref{f10.7extremejjas} and \ref{cri7extremejjas} lists the extreme rainfall events during the JJAS season. The first column represents the years of extreme of each solar activity feature (denoted as $y$), the second column the excess rainfall years, and the third column the deficient rainfall years. Values given in brackets indicate variation concerning solar extremes. 

\subsubsection{Relation of extreme rainfall years with Sunspot number}
\begin{figure*}[ht]
\centering
  \begin{subfigure}{0.5\textwidth}
    \includegraphics[width=\linewidth]{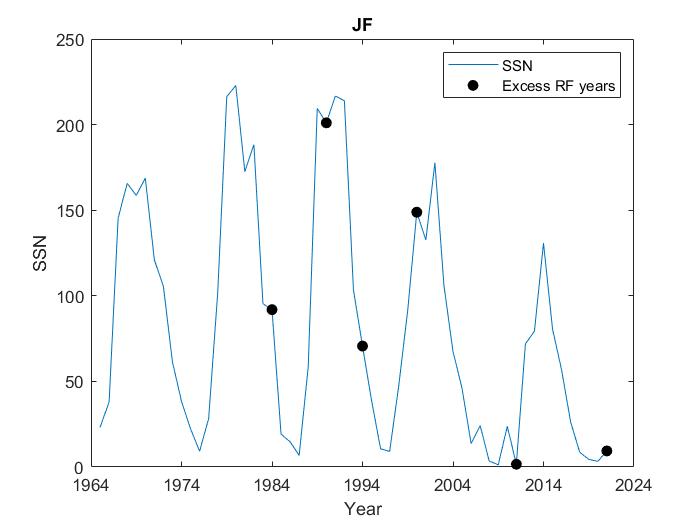}
     \label{fig:8a}
  \end{subfigure}%
  \begin{subfigure}{0.5\textwidth}
    \includegraphics[width=\linewidth]{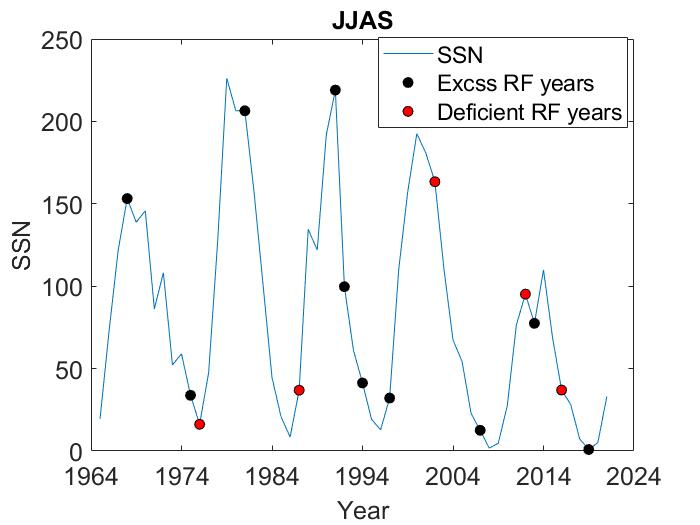}
    \label{fig:8b}
  \end{subfigure}  \\
\caption{The extreme years of JF and JJAS rainfall depicted on SSN plot}
\label{ssnexcess}
\end{figure*} 

The sunspot number SSN was considered first. Figure \ref{ssnexcess} plots the extreme rainfall years during the JF and JJAS seasons on the SSN curve. This figure shows slight variations in the maximum and minimum values of SSN during the JF and JJAS seasons. During the JF season, the SSN showed a maximum during 1970, 1980, 1991, 2002, and 2014 and a minimum during 1965, 1976, 1987, 1997, 2009, and 2020. In the JJAS season, maximum SSNs occurred in 1968, 1979, 1991, 2000, and 2014, and minimum SSNs occurred in 1965, 1976, 1986, 1996, 2008, and 2019. 
\begin{itemize}
   \item  {Excess rainfall studies} \\
 {The six excess rainfall years observed during the JF season are listed in Table \ref{ssnextremejf}.
Two occurred around $\pm$ 1 year, and the other two around $\pm$ 2 years of extreme solar activity (solar maximum/solar minimum). The remaining two excess rainfall years were observed further from the solar minimum, i.e., three years before the solar minimum.
Ten excess rainfall years were identified during the JJAS season and given in Table \ref{ssnextremejjas}. Three occurred at solar extremes, five around $\pm$1 years of solar extremes, and two around $\pm$2 years of solar extremes. }
   \item  {Deficient rainfall studies} \\
 {During the JF season, there were no years with deficient rainfall. Five deficient rainfall events were noted during the JJAS season and are listed in Table \ref{ssnextremejjas}. One occurred at the extreme of the solar cycle, another one a year later, and the remaining three occurred around $\pm$2 years of solar extreme.}
\end{itemize}

\begin{table*}[ht]
   \centering
  \caption{Extreme rainfall years along with extreme SSN years, during JF season}
     \label{ssnextremejf}
    \begin{tabular}{cc} \hline
       Years of extreme SSN (y) & Excess rainfall years \\ \hline
    1965 (min) & \\
       1970 (max) & \\
       1976 (min) &  \\
       1980 (max) &  \\
       1987 (min) & 1984 (y-3) \\
       1991 (max) & 1990 (y-1) \\
       1997 (min) & 1994 (y-3)\\
       2002 (max) & 2000 (y-2) \\
       2009 (min) & 2011 (y+2) \\
       2014 (max) & \\
       2020 (min) & 2021 (y+1) \\ \hline
  \end{tabular}
   \end{table*}

\begin{table*}[ht]
   \centering
  \caption{Extreme rainfall years along with extreme SSN years, during JJAS season}
     \label{ssnextremejjas}
    \begin{tabular}{ccc} \hline
       Years of extreme SSN (y) & Excess rainfall years & Deficient rainfall years \\ \hline
       1965 (min) & & \\
       1968 (max) & 1968 (y) & \\
       1976 (min) & 1975 (y-1) & 1976 (y) \\
       1979 (max) & 1981 (y+2) & \\
       1986 (min) & &  1987 (y+1)  \\
       1991 (max) & 1991 (y), 1992 (y+1) &  \\
       1996 (min) & 1994 (y-2), 1997 (y+1) & \\
       2000 (max) &  & 2002 (y+2) \\
       2008 (min) & 2007 (y-1) \\
       2014 (max) & 2013 (y-1) & 2012 (y-2), 2016 (y+2)\\
       2019 (min) & 2019 (y) & \\ \hline
  \end{tabular}
   \end{table*}
   
   \subsubsection{Relation of extreme rainfall years with F10.7 Index}

\begin{figure*}[ht]
\centering
  \begin{subfigure}{0.5\textwidth}
    \includegraphics[width=\linewidth]{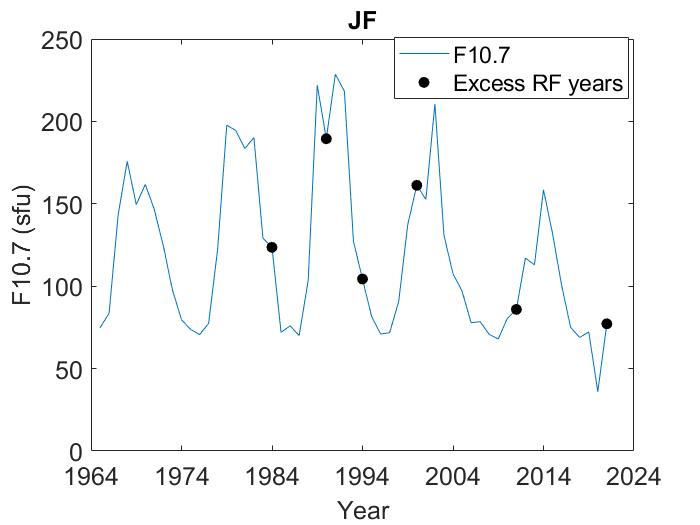}
     \label{fig:9a}
  \end{subfigure}%
  \begin{subfigure}{0.5\textwidth}
    \includegraphics[width=\linewidth]{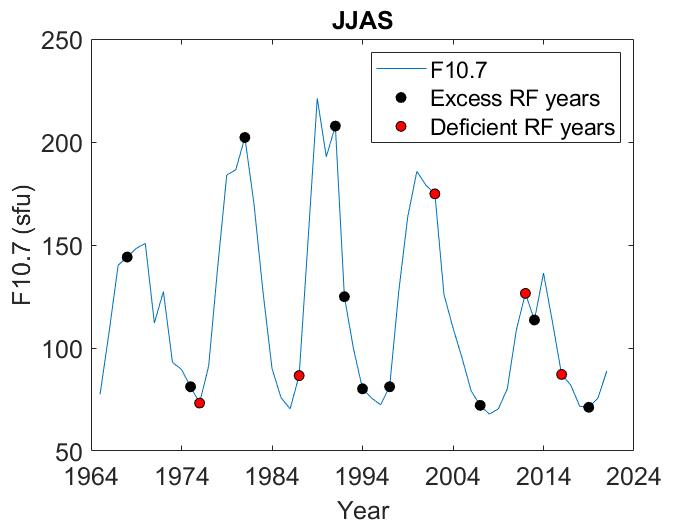}
     \label{fig:9b}
  \end{subfigure}  \\
\caption{The extreme years of (a) JF and (b) JJAS rainfall depicted on F10.7 plot}
\label{f107excess}
\end{figure*} 

The extreme rainfall years were then analyzed in terms of the F10.7 Index. Figure \ref{f107excess} shows the extreme rainfall years during the JF and JJAS seasons plotted on the F10.7 curve. The two seasons show different maximum and minimum values for each solar cycle. During the JF season, F10.7 showed peak values in 1968, 1979, 1991, 2002, and 2014 and trough values in 1965, 1976, 1987, 1996, 2009, and 2020. 
In the JJAS season, F10.7 revealed maximum values in 1970, 1981, 1989, 2000, and 2014 and minimum values in 1965, 1976, 1986, 1996, 2008, and 2019.

\begin{itemize}
    \item  {Excess rainfall studies} \\
     { During the JF season, six years of excess rainfall were noticed and are listed in Table \ref{f10.7extremejf}. It was noted that four out of these occurred around solar minimum and the remaining two around solar maximum. The two years of excess rainfall were around $\pm$1 years of solar extreme, and the three were around $\pm$2 years of solar extreme.
Details of the excess rainfall events during the JJAS season are listed in Table \ref{f10.7extremejjas}. Of the ten years of excess rainfall, two occurred at the solar extreme, four at around $\pm$1 years of solar extreme, and three around $\pm$2 years of solar extreme. One of them was visible a bit further from the solar extreme, i.e.,3 years after the solar maximum.}

    \item  {Deficient rainfall studies} \\
   { During the JJAS season, five years of deficient rainfall were observed, and details are given in Table \ref{f10.7extremejjas}. One of them occurred at solar extreme, i.e., at solar minimum. The remaining one is at a year after solar extreme, i.e., solar minimum, and the other two are around $\pm$2 years of solar extreme, i.e., solar maximum.}
\end{itemize}
   \begin{table*}[ht]
   \centering
  \caption{Extreme rainfall years along with extreme F10.7 years, during JF season}
     \label{f10.7extremejf}
    \begin{tabular}{cc} \hline
       Years of extreme F10.7 (y) & Excess rainfall years \\ \hline
       1965 (min) & \\
       1968 (max) & \\
       1976 (min) &  \\
       1979 (max) & \\
       1987 (min) & 1984 (y-3) \\
       1991 (max) & 1990 (y-1) \\
       1996 (min) & 1994 (y-2)\\
       2002 (max) & 2000 (y-2) \\
       2009 (min) & 2011 (y+2) \\
       2014 (max) & \\
       2020 (min) & 2021 (y+1) \\ \hline
  \end{tabular}
   \end{table*}

   \begin{table*}[ht]
   \centering
  \caption{Extreme rainfall years along with extreme F10.7 years, during JJAS season}
     \label{f10.7extremejjas}
    \begin{tabular}{ccc} \hline
       Years of extreme F10.7 (y) & Excess rainfall years & Deficient rainfall years \\ \hline
       1965 (min) & & \\
       1970 (max) & 1968 (y-2) & \\
       1976 (min) & 1975 (y-1) & 1976 (y) \\
       1981 (max) & 1981 (y) & \\
       1986 (min) & & 1987 (y+1) \\
       1989 (max) & 1991 (y+2), 1992 (y+3) & \\
       1996 (min) & 1994 (y-2), 1997 (y+1) & \\
       2000 (max) &  & 2002 (y+2) \\
       2008 (min) & 2007 (y-1) \\  
       2014 (max) &  2013 (y-1) & 2012 (y-2), 2016 (y+2)\\
       2019 (min) & 2019 (y) & \\ \hline
  \end{tabular}
   \end{table*}

   \subsubsection{Relation of extreme rainfall years with Cosmic ray intensity}

\begin{figure*}[ht]
\centering
  \begin{subfigure}{0.5\textwidth}
    \includegraphics[width=\linewidth]{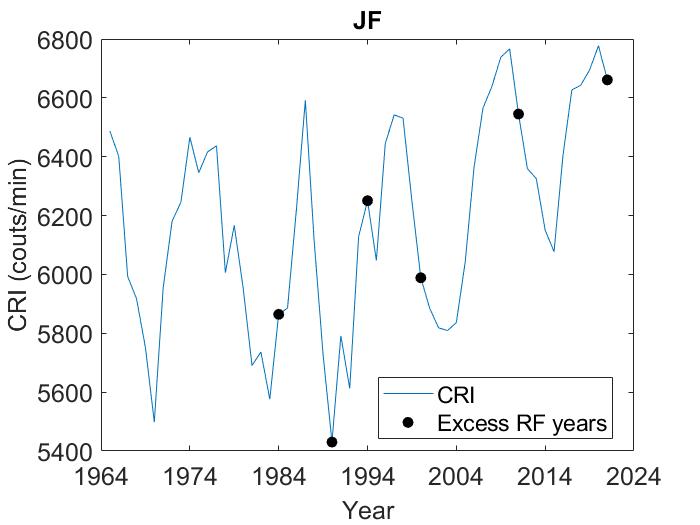}
     \label{fig:10a}
  \end{subfigure}%
  \begin{subfigure}{0.5\textwidth}
    \includegraphics[width=\linewidth]{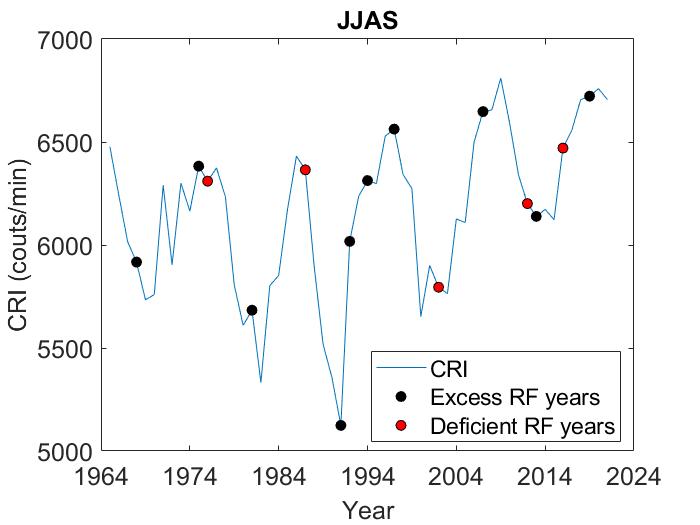}
     \label{fig:10b}
  \end{subfigure}  \\
\caption{The extreme years of (a) JF and (b) JJAS rainfall depicted on CRI plot}
\label{criexcess}
\end{figure*}

Lastly, the relationship of extreme rainfall years with CRI was studied. Figure \ref{criexcess} depicts the extreme rainfall years during the JF and JJAS seasons plotted on CRI curves. {During the JF season, CRI displayed minimum values during 1970, 1983, 1990, 2003, and 2015 and maximum values during 1965, 1974, 1987, 1997, 2010, and 2020.} 

 {In the JJAS season, CRI showed minimum values in 1969, 1982, 1991, 2000, and 2015 and maximum values in 1965, 1975, 1986, 1997, 2009, and 2020.}

\begin{itemize}
 \item  {Excess rainfall studies} \\
 {A list of the extreme rainfall years during the JF and JJAS seasons is given in Tables \ref{criextremejf} and \ref{cri7extremejjas}, respectively. Out of the six years of excess rainfall during the JF season, one was observed at solar extreme, three at around $\pm$1 years of solar extreme, and the rest further away around $\pm$3 years of solar extreme.
During the JJAS season, three years of excess rainfall occurred at the solar extreme, four at around $\pm$1 years of solar extreme, and the rest at around $\pm$2 years of solar extreme.}
    \item {Deficient rainfall studies} \\
    Five years of deficient rainfall was observed in the JJAS season, given in Table \ref{cri7extremejjas}. Out of them, three were observed at around $\pm$1 years of solar extreme, one a year after solar extreme, i.e., solar minimum, and one three years after solar extreme, i.e., solar minimum.
\end{itemize}

\begin{table*}[ht]
   \centering
  \caption{Extreme rainfall years along with extreme CRI years, during JF season}
     \label{criextremejf}
    \begin{tabular}{cc} \hline
       Years of extreme CRI (y) & Excess rainfall years \\ \hline
       1965 (max) & \\
       1970 (min) & \\
       1974 (max) &  \\
       1983 (min) & 1984 (y+1) \\
       1987 (max) &  \\
       1990 (min) & 1990 (y) \\
       1997 (max) & 1994 (y-3) \\
       2003 (min) & 2000 (y-3) \\
       2010 (max) & 2011 (y+1) \\
       2015 (min) & \\
       2020 (max) & 2021 (y+1) \\ \hline
  \end{tabular}
   \end{table*}

\begin{table*}[ht]
   \centering
  \caption{Extreme rainfall years along with extreme CRI years, during JJAS season}
     \label{cri7extremejjas}
    \begin{tabular}{ccc} \hline
       Years of extreme CRI (y) & Excess rainfall years & Deficient rainfall years \\ \hline
       1965 (max) & & \\
       1969 (min) & 1968 (y-1) & \\
       1975 (max) & 1975 (y) &  1976 (y+1) \\
       1982 (min) & 1981 (y-1) & \\
       1986 (max) & & 1987 (y+1)\\
       1991 (min) & 1991 (y), 1992 (y+1) & \\
       1997 (max) & 1994 (y-3), 1997 (y) & \\
       2000 (min) &  & 2002 (y+2) \\
       2009 (max) & 2007 (y-2)  &  2012 (y+3)\\
       2015 (min) & 2013 (y-2) & 2016 (y+1)\\
       2020 (max) & 2019 (y-1) & \\ \hline
  \end{tabular}
   \end{table*}

   {When looking at the results related to sunspot number, F10.7 Index, and cosmic ray intensity, we could conclude that Kerala experienced extreme rainfall, mostly when solar activity was at its extreme, i.e., when solar maximum or minimum occurred or within $\pm$2 years of solar extremes. It is possible to predict extreme rainfall events in Kerala by understanding the extreme variability of solar activity. As the rainfall deficiency/excess was observed almost equally around both extremes, it was impossible to determine whether extreme events were more prevalent during solar maximum or minimum.}
   
   According to the SSN and F10.7 analyses, excessive rainfall was more prevalent during solar minimum during the JF season and solar maximum during the JJAS season. The deficient rainfall events were noted more around solar maximum according to the SSN and F10.7 results. 

The results obtained are in agreement with earlier studies carried out in India. \cite{bhalme1981cyclic} reported that the Flood Area Index over India was associated with the double sunspot cycle. 
During alternate solar cycles, \cite{Ananthakrishnan1984} observed significantly more excess rainfall years during the ascending phase.
 \cite{jain1997correlation} considered the Udaipur subtropical region in Rajasthan to check any possible relation between solar activity and its rainfall and noted that the periodicity of floods and droughts are well correlated with sunspot main periods and/or quasi-periods. \cite{Bhattacharyya2005} revealed that high rainfall is linked with high solar activity and low rainfall with low solar activity. \cite{Azad2011} while studying the relation of extreme Indian monsoon rainfall over sub-divisions from west central and peninsular India with sunspots, reported that the maxima of even sunspot cycles coincided with excess rainfall (with +1 year error) and the minima of odd sunspot cycles coincided with deficit rainfall (with $\pm$2 year error). 

Several studies have reported how solar activity affects extreme rainfall events worldwide. In the United States, it was observed that the drought cycle is related to the double (Hale) sunspot cycle \citep{mitchell1979,cook1997}. \cite{Vaquero2004SolarSI} evaluated the number of floods recorded for the Tagus river basin, Central Spain, and it was noted that the probability of floods increased during the episodes of high solar activity. A study on the levels of Lake Victoria, East Africa, revealed the influence of solar activity on the levels through rainfall. The rainfall maxima had a lagged relationship with the sunspot maxima by one year, leading to the lake level maxima \citep{Stager2007}. Sunspot number directly correlated with the flood/drought of the Second Songhua river basin, China, and flood years appeared in Solar Maximum Year, years after Solar Maximum Year, and Solar Minimum Year \citep{hong2015}. Studies relating the response of extreme precipitation to solar activity in typical regions of the Loess Plateau in Yan'an, China, observed that the maximum precipitation occurred mainly during solar maximum and was correlated \citep{Li2017}. \cite{YU2019} also reported that droughts and floods in the Southern Chinese Loess Plateau were synchronous with solar activities, at least on decadal timescales. 

There are instances of opposing results being reported as well. \cite{Wirth2013} found that flood frequency in the European Alps increased during cool periods, which coincided with low solar activity. In studies relating to River Ammer floods in Germany, \cite{Rimbu2021} observed that the frequency of flood years is relatively high with low solar activity and vice versa. \cite{Li2023} investigated the time-lagged correlations between solar activity and summer precipitation in the mid-lower reaches of the Yangtze River, China, and it was observed that the sunspot number has a negative correlation with precipitation, with a time lag of 11 months.

Solar activity is observed to influence global rainfall patterns through complex mechanisms \citep{Wang2012, Li2023}. The variation in solar irradiance reaching the Earth during increased solar activity can influence atmospheric circulation patterns and thereby the distribution and intensity of rainfall \citep{Kodera2007, Nazari-Sharabian2020}. Total solar irradiance (TSI) may affect sea surface temperature, which changes atmospheric circulation and affects rainfall patterns \citep{Soon1996}. Furthermore, solar activity influences levels of ultraviolet radiation, which may impact the Earth's ozone layer and the temperature gradients in the stratosphere. This, in turn, can alter atmospheric circulation patterns \citep{Baldwin2005,rycroft}.
Similar periods in the time series of rainfall and solar activity often implied a possible relation between them \citep{Nitka2019, HEREDIA2019105094}. Solar activity has a global impact along with other factors like geographical location, local climate system, atmospheric circulation, feedback mechanisms, etc, influencing precipitation diversely \citep{Brunner2021}.
Solar activity affects the production of cloud condensation nuclei \citep{Svensmark2007,svensmark2019force} and, eventually, affects rainfall distribution. Various natural elements like the El Niño/Southern Oscillation (ENSO) and the North Atlantic Oscillation (NAO) are important indices that significantly affect rainfall patterns and are also influenced by solar activity \citep{Leamon2021,kuroda}. The relationship between solar activity and climate is intriguing and complex. Further research is necessary to understand this association thoroughly and to reach well-supported conclusions.

\section{Conclusions} \label{conclusions}
We have evaluated the possible impact of solar activity on extreme rainfall events over 57 years in Kerala, India. Different solar indices, i.e., Sunspot number, F10.7 Index, and cosmic ray intensity, were used, and their variation with rainfall was studied. Correlation studies were performed for each solar cycle, starting with Solar Cycle 20. It was observed that rainfall in Kerala is correlated with sunspot activity, with varying significance. The SSN showed the strongest correlation with rainfall among the three solar activity indices: SSN, F10.7, and CRI. A significant correlation exists between rainfall and solar activity during the winter and monsoon seasons of Solar Cycle 21.
In comparison to the monsoon season, the winter season revealed stronger solar influences on its rainfall. The years when rainfall in Kerala was excess or deficient were identified, and its connection with the different solar indices was studied. Most years with excess or deficient rainfall have been observed to coincide with solar extremes, such as solar maximum or minimum (within $\pm$2 years). It is possible to predict extreme rainfall events in Kerala based on the coincidence and reduce their disastrous impacts. Gaining more profound knowledge about solar-climate interactions helps predict climate changes and improve the existing climate models, leading to environmental sustainability \citep{Gautam2024}.

\section*{Acknowledgements}
First author acknowledges the financial assistance from the University Grants Commission (UGC), India, under Savitribai Jyotirao Phule Fellowship for Single Girl Child (SJSGC) (F. No. 82-7/2022(SA-III) dated 07/02/2023). Second author acknowledges the financial assistance from Department of Science and Technology (DST), Ministry of Science and Technology, India under INSPIRE Fellowship (Award Letter No. IF180235 dated 08/02/2019).

\vspace{-1em}




\end{document}